# Open Boundary Simulations of Proteins and their Hydration Shells by Hamiltonian Adaptive Resolution Scheme

*Thomas Tarenzi[†‡], Vania Calandrini\*[§], Raffaello Potestio[⊥], Alejandro Giorgetti[§#], Paolo Carloni[‡§]*

[†]Computation-Based Science and Technology Research Center CaSToRC, The Cyprus Institute, 20 Konstantinou Kavafi Street, 2121, Aglantzia, Nicosia, Cyprus.

[‡]Department of Physics, Faculty of Mathematics, Computer Science and Natural Sciences, Aachen University, Otto-Blumenthal-Straße, 52074 Aachen, Germany.

[§]Computational Biomedicine, Institute for Advanced Simulation IAS-5, and Institute of Neuroscience and Medicine INM-9, Forschungszentrum Jülich, 52425 Jülich, Germany.

[⊥]Max Planck Institute for Polymer Research, Ackermannweg 10, 55128 Mainz, Germany.

[#]Department of Biotechnology, University of Verona, Ca' Vignal 1, Strada Le Grazie 15, 37134 Verona, Italy.

# ABSTRACT

The recently proposed Hamiltonian Adaptive Resolution Scheme (H-AdResS) allows to perform molecular simulations in an open boundary framework. It allows to change on the fly the resolution of specific subset of molecules (usually the solvent), which are free to diffuse between the atomistic region and the coarse-grained reservoir. So far, the method has been successfully applied to pure liquids. Coupling the H-AdResS methodology to hybrid models of proteins, such as the Molecular Mechanics/Coarse-Grained (MM/CG) scheme, is a promising approach for rigorous calculations of ligand binding free energies in low-resolution protein models. Towards this goal, here we apply for the first time H-AdResS to two atomistic proteins in dual-resolution solvent, proving its ability to reproduce structural and dynamic properties of both the proteins and the solvent, as obtained from atomistic simulations.

## 1. INTRODUCTION

Ligand docking is a fundamental tool in current pharmaceutical design [1-5]. However, standard docking and molecular simulation-based approaches [6-13] may face severe challenges [14-16] in predicting accurate binding poses and ligand affinities in the case of membrane proteins (which constitute as much as 60% of the overall proteins targeted by FDA-approved drugs [17]), because of the lack of experimental structural information [18-19] and the low sequence identity (SI, often below 20%) between the members of some large families [17, 18] (such as G-protein-coupled receptors, GPCRs). In these systems the inaccuracy in the modeled side chains orientation may hamper the correct description of protein/ligand interactions [20]. Keeping this pharmacological application in mind, our team developed an hybrid Molecular Mechanics/Coarse-Grained (MM/CG) approach [21] specifically conceived to predict ligand poses in membrane proteins' low resolution models[22]. Within this scheme the binding cavity, fully solvated, is represented with atomistic detail (MM region), while the rest of the protein is represented with a coarse grained resolution (CG region) (Figure 1). An intermediate region is defined between the MM and CG parts in order to smoothly couple the two levels of resolution. This scheme allows the description of the atomistic details of the ligand/receptor binding without introducing unnecessary bias coming from the potentially wrong orientations of the side chains located far from the binding site.

The approach turned out to be able to reproduce the main features of membrane proteins' X-ray structures[21] and to predict, consistently with experiments, the binding poses of ligands binding to two targets with extremely low SI with their templates (namely bitter taste receptors TAS2R38[23] and TAS2R46[20], SI<15%). However, a limitation in the applicability of the method, particularly in predicting ligand binding free energies, is that the boundary potential used to prevent the evaporation of the water droplet solvating the binding cavity[21] might lead to artifacts in the solvent dynamics. The key role of water for drugs binding to their target membrane receptors have been widely recognized [24-26], ranging from ligand discrimination[27-28], to determination of protein loops conformation [29-31], to mediation of ligand-receptor interactions[32-34].

Solvating completely the system with explicit water is the most straightforward approach to address this issue within the MM/CG approach. However, this increases the number of atoms and, consequently, the computational cost. In addition, the technical complexity in the definition of the interactions between the atomistic water molecules and the coarse-grained parts of the receptor might overcome the advantages provided by the dual-resolution model of the receptor.

Implementing a fully Hamiltonian dual resolution modeling for the solvent, in which the water molecules are described explicitly in the MM region (e.g. the water hemisphere in Figure 1) and coarse-grained outside, would be a physically sound and more elegant alternative (Figure 2). One of us (RP) and coworkers has recently developed one of such schemes, the so-called "Hamiltonian Adaptive Resolution Simulation scheme" (H-AdResS) [35-36]. Here, the interface between these two subdomains consists of a hybrid resolution region where water molecules change on the fly their resolution; they freely diffuse in and out of the MM region, thus removing possible artifacts introduced by a confining potential. H-AdResS interpolates the Hamiltonians of the two regions (as opposed to the force-interpolation scheme [37-38]) to couple smoothly the atomistic and the coarse-grained potentials at the interface. In this way, the H-AdResS scheme preserves by construction the Hamiltonian framework [35], generating a well defined statistical ensemble – the grand canonical one- for the region of interest, the MM region, provided that the CG reservoir is sufficiently large. This condition is easily met thanks to the efficient representation of molecules in the low-resolution subdomain. This setup allows rigorous ligand binding free energy calculations and affinity predictions [39]. First steps in this direction have been undertaken through the application of the force-based AdResS scheme to the solvation of a protein[40], specifically lysozyme, itself treated with a dual-resolution model. Here, only the residues constituting the enzyme's active site and the bound substrate were described at the atomistic level, while the remainder of the protein was modeled as an elastic network of beads (corresponding to the $C_\alpha$ atoms) connected by harmonic springs. This simulation approach, similar in spirit to the MM/CG one, was employed concurrently with an adaptive resolution representation of the solvent, where water molecules were atomistic only in proximity of the active site and coarse-grained elsewhere. Another step toward free energy calculations has recently been taken with the calculation of solvation free energies of amino acid side chain analogues[41], using the same methodology in the description of the solvent. In both works, the AdResS scheme was employed, which relies on the interpolation of different models at the level of forces and does not admit an energy-based formulation[38]. The H-AdResS method, which in contrast directly couples the potential energies of the models at different resolution, has never been applied to proteins but only to pure liquids. To establish the accuracy of this scheme for real-life biophysical systems, and as a first step towards the application to membrane proteins within the MM/CG scheme, we have used it to perform grand canonical-like simulations of cytosolic proteins in solution. Here, the proteins and the solvent in their proximity are treated at atomistic level, whilst the rest of the solvent, is treated at CG level. The atomistic region is therefore simulated in a grand canonical fashion, while the coarse-grained one plays the role of a particle reservoir. Both structural and dynamical properties of the proteins and of the solvent in the atomistic region are calculated, and the satisfactory comparison with fully atomistic simulations establishes the reliability of the method. Hence, this work paves the way towards the implementation of the H-AdResS scheme for a dual-resolution representation of the solvent coupled to the MM/CG description of membrane protein/ligand complexes, in order to accurately predict binding poses and ligand affinities. This implementation is currently ongoing.

This paper is organized as follows. After a theory section on the H-AdReS applied to biomolecular systems (Section 2), we present the first H-AdResS application to cytoplasmic proteins, providing details of the simulations performed (Section 3) and presenting the results in comparison with fully atomistic reference simulations (Section 4).

## 2. THE H-ADRESS SCHEME

Here we use the H-AdResS scheme [35] to describe a dual-resolution solvation water. The protein is treated at full atomistic level. Within this scheme, the Hamiltonian of the system reads:

$$H = K + V^{p/p-w} + \sum_{\alpha}^{N} \left\{ \lambda_\alpha V_\alpha^{MM} + (1-\lambda_\alpha) V_\alpha^{CG} \right\} - \sum_{\alpha}^{N} \Delta H(\lambda_\alpha). \quad (1)$$

In Eq. (1), $K$ is the kinetic energy of the system. $V^{p/p-w}$ represents the all-atom bonded and non-bonded interactions internal to the protein and the all-atom non-bonded protein-water interactions. Notice that in the previous applications of the method[35, 42], the term $V^{p/p-w}$ reduced to the potential describing the internal interactions of each solvent molecule. In the current application, solvent molecules are described by a rigid model, hence this contribution is discarded, but interactions internal to the protein and between the protein and the solvent are taken into account. The dual resolution H-AdResS scheme is applied only to the solvent. The weighted sum of $V_\alpha^{MM}$ and $V_\alpha^{CG}$ represents the hybrid non-bonded potential energy contribution of each water molecule $\alpha$. Specifically, $V_\alpha^{MM}$ and $V_\alpha^{CG}$ refer to the MM and CG contributions respectively[1], and $\lambda_\alpha$ is a smooth transition function of the CoM coordinates $\mathbf{R}_\alpha$ of water molecule α, $\lambda(\mathbf{R}_\alpha)$. It ranges from 1 in the MM region to 0 in the CG region. It couples the two levels of resolution depending on the water molecule's CoM position. This coupling allows each water molecule to change on the fly its resolution according to its CoM position. The sum is over the total number of water molecules, $N$. The water MM potential, $V^{MM}$ is the SPC/E force field [43]. $V^{CG}$ is derived from an independent all-atom simulation on pure water through iterative

---

[1] CG refers to a coarse-grain description of water molecules, where CG water molecules are represented as beads, centered on their center of mass (CoM). MM, instead, refers here to their all-atom description.

Boltzmann inversion [44]. Specifically, a mapping of a water molecule to one CG bead (corresponding to the water CoM) is used. This coarse-graining procedure was carried out using the VOTCA package [45]. The last term in Eq. (1) avoids a net flux of water, thus ensuring a flat density profile across the regions at different resolution. This "correction term" reads:

$$\Delta H(\lambda_\alpha) = \frac{\Delta F(\lambda_\alpha)}{N} + \frac{\Delta p(\lambda_\alpha)}{\rho_0} \equiv \Delta \mu(\lambda_\alpha) = \frac{\Delta G(\lambda_\alpha)}{N}. \quad (2)$$

$\Delta F(\lambda_\alpha)$, $\Delta p(\lambda_\alpha)$ and $\Delta G(\lambda_\alpha)$ are, respectively, the Helmholtz free energy difference, the pressure difference, and the Gibbs free energy difference between a system with hybrid Hamiltonian at resolution $\lambda_\alpha$ and the CG system with $\lambda=0$. $\rho_0$ is the reference (target) density of water. The first term compensates on average the so-called drift force arising from the coupling of the CG and MM Hamiltonians. The second term compensates the so-called thermodynamic force, arising from the pressure imbalance between the CG and MM subsystems (supporting information SI 1 reports the derivation of these terms). We note that the sum of the compensation terms corresponds to the change in chemical potential $\mu$ [35]. This shows that the H-AdResS formulation not only provides a uniform solvent density throughout the simulation box, but also a constant chemical potential $\mu$. Hence, it generates a grand canonical ensemble in the high-resolution (atomistic) region: the number of water molecules in the atomistic region can vary, while the chemical potential is kept constant thanks to the application of the Gibbs free energy correction.

### 3. METHODS

**Systems set-up.** The systems selected as test cases were the human metallochaperone atox1 in the apo form (PDB code 1TL5)[46] and the human cyclophilin J (PDB code 2OJU). These are two soluble, globular and structured proteins, remarkably different in size (68 and 166 residues, respectively) and gyration radius (1.1 nm and 1.4 nm, respectively). The latter are used as test-case in order to find a general criterion for the choice of the optimal atomistic size, $d_{at}$, based on an intrinsic property of the simulated system, such as the gyration radius of the protein.

Both systems were solvated in a water box. The final box edge after equilibration was ~10.9 nm for atox1 and ~11.9 nm for cyclophilin. In order to compare H-AdResS and fully atomistic systems, each system underwent fully atomistic and H-AdResS-based simulations. The proteins were described using the AMBER99SB-ILDN force field [47], and were treated with fully atomistic details during the whole simulations. In the H-AdResS based simulations, the atomistic region had a spherical geometry, with radius $d_{at}$ ranging from 2.0 to 3.0 nm in the case of atox1, and from 2.5 to 3.5 nm in the case of cyclophilin J. The atomistic region was centered on the center of the simulation box, corresponding also to the CoM of the protein. In all cases, the thickness of the hybrid region was 1.2 nm.

**MD simulations.** Both systems underwent ~400 ns long equilibration runs, which were carried out in NPT ensemble. The integrator used was a leap frog stochastic dynamics integrator [48], which allows stochastic temperature-coupling, by adding to the Newton's equation of motion a friction and a noise term. The temperature was set to 300 K and the inverse friction constant to 0.2 ps. In addition, a Berendsen barostat was used, with pressure set to 1 atm and time constant to 0.5 ps. A time integration step of 2 fs was used during the first 200 ns of the equilibration runs, constraining the hydrogen-containing bonds by LINCS algorithm [49]. The last 200 ns of the equilibration were performed using a time integration step of 0.5 fs without any constraint on the hydrogen-containing bonds. Starting from the previously equilibrated systems, the production runs were carried out using the same integrator as before, with an integration time step of 0.5 fs, without any constraint on the protein bonds containing hydrogen atoms. The length and sampling time step of the production runs used to calculate the structural and dynamical properties of both water and protein are reported in Table 1. Electrostatics was treated using the reaction field method [50]. A cutoff of 1.2 nm was used for all non-bonded interactions. All simulations were performed with periodic boundary conditions and the minimum image convention. The protein center of mass was kept fixed in the center of the box during the simulation.

Equilibration runs and all-atoms simulations were done using GROMACS 5.1.2 [51], while an in-house version of GROMACS 4.6 [52] implementing H-AdResS was used to run the H-AdResS-based simulations.

**Post-processing analyses.** Data analysis was performed using GROMACS utilities, the web server VADAR[53] and in-house scripts. Molecular images and visual inspections of the systems were made with VMD 1.9.2 [54]. In order to compare H-AdResS-based and fully atomistic systems, analyses were performed on molecules in the atomistic part of the H-AdResS simulations and in a sphere of the all-atom systems with the same center and radius. Given the impact of the protein on water structure and dynamics [55-56], especially in the hydration layers, several of the calculations were limited to subpopulations of water, namely, the first hydration shell and the bulk atomistic water. The former was defined as consisting of all water molecules whose oxygen atom was within a cutoff distance of 0.7 nm from the nearest protein heavy atom. The bulk atomistic water region was defined by all water molecules located more than 0.7 nm apart from the protein surface and less than $d_{at}$ nm from the center of the atomistic region. The maintenance of the folded state of the proteins has been assessed with the program DSSP [57], computing secondary structure for each residue during the whole simulations. To evaluate the protein stability and its interaction with the solvent, we computed also the solvent accessible surface area (SASA) per residue [58]. Furthermore, to assess any difference in the protein flexibility, we calculated the root mean square fluctuations (RMSF) of the $C_\alpha$ positions in the trajectory. In order to quantify the plasticity of proteins' backbone residues, we computed the Protein Angular Dispersion of the angle $\omega$ (PAD$_\omega$)[59], where $\omega$ is the sum of the dihedral angles $\varphi$ and $\psi$. PAD$_\omega$, defined as in ref. [59], can range between 0° (when $\varphi$ and $\psi$ dihedral angles of the residue do not change across the protein structures along the trajectory) and 180° (when the two dihedral angles assume random values). For the atox1 system, an analysis of the average dihedral angles $\varphi$, $\psi$ and $\chi$ and of the main-chain hydrogen bonds was performed using the web server VADAR[53]. The average values and their standard deviation were calculated over 15 snapshots along the simulations. The proteins dynamics has been assessed by computing the THz IR spectra with VMD 1.9.2 [54]; the frequency spectrum is computed as the Fourier transform of the time correlation function of the total dipole moment $\mathbf{M}(t)$, applying to the plain power spectrum a prefactor dependent on the temperature:

$$\alpha(\omega) = \frac{\omega^2 \hbar}{2\pi kT} \int_{-\infty}^{+\infty} d\exp(-i\omega t)\langle \mathbf{M}(0)\mathbf{M}(t)\rangle \qquad (3)$$

where:

$$\mathbf{M}(t) = \sum_\alpha q_\alpha \mathbf{R}_\alpha(t) - \left\langle \sum_\alpha q_\alpha \mathbf{R}_\alpha \right\rangle \qquad (4)$$

Regarding solvent's properties, the water density as a function of the distance from the center of the atomistic region has been computed. A more local description of the solvent structure is given by the water oxygen-oxygen and oxygen-hydrogen radial distribution functions, and by the probability distribution of the tetrahedral order parameter [60] for water molecules in the hydration shell and for water molecules in the bulk atomistic region. The latter quantifies to which extent a given water molecule and its four nearest neighbors adopt a tetrahedral arrangement.

To probe water's dynamics, we computed the mean square displacement (MSD) to collect information on the diffusive motion of atomistic water:

$$MSD = \langle (\mathbf{r}-\mathbf{r}_0)^2 \rangle = \frac{1}{N}\sum_{n=1}^{N}(\mathbf{r}_n(t)-\mathbf{r}_n(0))^2 \qquad (5)$$

where $\mathbf{r}$ is the position of the water oxygen atoms and $N$ is the number of atoms to be averaged.

When varying the size of the atomistic region to probe its influence on protein and hydration water behaviors, we estimated **(i)** the protein's radius of gyration:

$$R_g = \left(\frac{\sum_i \|\mathbf{r}_i\|^2 m_i}{\sum_i m_i}\right)^{\frac{1}{2}} \qquad (6)$$

where $m_i$ is the mass of atom $i$, and $\mathbf{r}_i$ is its position with respect to the center of mass of the molecule; **(ii)** the number of intra-protein hydrogen bonds, relative to the number in fully atomistic simulations; **(iii)** the error in the RMSF for the protein $C_\alpha$ atoms, relative to the RMSF in the fully atomistic system:

$$\left\langle \frac{\delta x}{x} \right\rangle = \frac{1}{N}\sum_{i=1}^{N}\frac{|x_{i,A}-x_{i,H}|}{x_{i,A}} \qquad (7)$$

where $x_{i,A}$ is the RMSF for the $C_\alpha$ of residue $i$ in the fully atomistic system, $x_{i,H}$ is the corresponding value in the H-AdResS system, and the sum is over all $N$ residues of the protein. Protein's dynamics has been investigated by computing IR spectra, as in Eq. (3). Going back to the water properties, after having checked the water density profiles throughout the simulation box, we probed hydration water's structure by computing the tetrahedral order parameter, and its dynamics by computing the reorientation time correlation function (tcf). The latter is defined as $\langle P_2[\mathbf{u}(0)\cdot\mathbf{u}(t)]\rangle$, where $P_2$ is the second-order Legendre polynomial and $\mathbf{u}$ is the vector along the water OH bond, for molecules in the hydration shell at the time origin. The characteristic reorientation times were computed as the integral of the corresponding tcfs.

## 4. RESULTS AND DISCUSSION

The aim of the following analysis is to demonstrate the applicability of the H-AdResS method also to water solvating complex biological systems, by performing comparison between structural and dynamical properties of both proteins and solvent, computed in the H-AdResS and fully atomistic simulations.

We focus on the atox1 protein (also called HAH1 or Atx1), and the cyclophilin J proteins.

The first is a copper chaperone [61-62]. It delivers copper(I) to both the Menkes and the Wilson diseases correlated proteins, membrane-bound ATPases. These translocate copper in the trans-Golgi network or across the plasma membrane [63-65]. Atox1 presents a ferrodoxin-like βαββαβ fold (Figure 3.a) [46].

Cyclophilin J is a member of the peptidyl-prolyl *cis-trans*-isomerase superfamily [66]. Cyclophilins have been implicated in many pathological processes, including virus infections [67], rheumatoid arthritis [68], cardiovascular diseases [69] and cancer [70-71]. In particular, Cyclophilin J has been suggested to play a determinant role in hepatocellular carcinoma carcinogenesis [66]. The protein contains helices and one β-barrel (Figure 3.b), composed of eight antiparallel β-strands arranged clockwise [72].

We report hereon structure, conformational fluctuations and dynamics of proteins (i) and water (ii) in the atomistic region of H-AdResS simulations (Figure 3.c) and in the corresponding region of fully atomistic simulations. A key parameter of the simulation is the radius of the atomistic region ($d_{at}$), defined as the radius of the sphere centered on the center of geometry of the protein. This dictates the amount of all-atom water molecules around the protein. The longer $d_{at}$, the more water molecules will surround the biomolecule. Here, $d_{at}$ is set to 3.0 nm and 3.5 nm for atox1 and cyclophilin J, respectively. We close the section with an investigation of the influence of $d_{at}$ on the calculated properties. We therefore run 10 ns H-AdResS simulations with reduced values of $d_{at}$, namely 2.5, 2.0 nm for atox1, and 3.0, 2.5 nm for cyclophilin. These values are still large enough to avoid interactions between the atomistic protein and the fully coarse-grained water molecules, which are not defined in our simulations. Diminishing further $d_{at}$, the CG solvent would fall within the interaction cutoff of some protein atoms, leading to unphysical situations. This discussion may help the reader interested in running H-AdResS simulations to determine the optimum system size, on the basis of the accuracy of the system's physical description.

**Proteins: structure and dynamics.** The root mean square deviation (RMSD) of backbone atoms during equilibration phase plotted as a function of time shows that the system is well equilibrated (Figure SI 2). Focusing on the production simulations, the secondary structure (Figure SI 3) appears to be fully maintained during the dynamics, showing agreement between the H-AdResS and atomistic simulations. The solvent accessible surface area (SASA) per residue (Figure SI 4) displays fair agreement with the fully atomistic reference. However, some discrepancies in the SASA plots are encountered for some residues exposed to the solvent. The residual plots are shown in Figure SI 4; the maximum residual corresponds to a difference of 11% with respect to the fully atomistic reference. This may be ascribed to the small differences in the water density in the two simulations (see below). The root mean square fluctuation (RMSF) of the $C_\alpha$ positions in the trajectory (Figure 4) and the plasticity of proteins' backbone, quantified by the parameter $PAD_\omega$ (Figure 5), compare well with the fully atomistic simulation. Therefore, we conclude that the multiscale simulations reproduce structure and conformational fluctuations of the all-atom MD with good accuracy. Additional analyses have been conducted on the atox1 system, for which published results from atomistic simulations are available[73]. The dihedral angles $\psi$ and $\varphi$, averaged over different snapshots along the trajectories, are within the error bars for most of the residues (Figure 6); the discrepancies between the results published by *Calandrini et al*[73]. and those from our atomistic and H-AdResS simulations might lie in the different force field used to describe the protein (AMBER99SB-ILDN force field[47] in our simulations, and AMBER parm99 force field[74] with the "Stony Brook" modification[75] for the backbone torsions in *Calandrini et al.*[73]). The average number of residues with Gauche+ torsional angle $\chi$, Gauche- $\chi$, and trans $\chi$, and the average values of these angles, are reported in Table 2. They show a very good agreement among the three simulations. Finally, distances, angles and energies of the main-chain hydrogen bonds in atox1 are reported in Figure 7. All the results lie within the error bars. These results provide an estimation of the dispersion of the values as obtained from two fully independent atomistic simulations, and the compatibility of our hybrid simulations with the atomistic ones within the dispersion values, further corroborates the reliability of H-AdResS in reproducing atomistic simulations.

We computed the IR spectra of the proteins ranging from 0 to ~200 cm$^{-1}$ (Figure 8). This region is sensitive to the functionally relevant, global and sub-global collective modes of the protein [76]. Our H-AdResS simulations reproduce the spectrum of the fluctuations of the permanent dipole moment with respect to the fully atomistic case.

**Water structure.** The calculated water density profiles throughout the simulation box (Figure 9) are in fair agreement with those in the fully atomistic simulations, apart from small deviations (quantified in the residual plots, displayed in the same figure). While some of these are present also in the atomistic region, the highest discrepancies are observed in the hybrid region, where they correspond to the 5% of the reference density value. Although these deviations might be reflected in the SASA, the other structural/dynamical properties analyzed do not seem to be affected. Notice that the depletion of water molecules in proximity to the center of the atomistic region is attributable to the presence of the protein; therefore, these graphs establish that the (all-atoms) proteins are entirely confined in the atomistic region of the solvent.

To provide a description of the solvent structure, we calculated the water oxygen-oxygen and oxygen-hydrogen radial distribution functions (rdf's, Figure SI 5), along with the probability distributions of the tetrahedral order parameter for water molecules [60] (Figure SI 6), computed for water molecules in the first hydration shell and in the bulk atomistic region. The results agree well with all-atom simulations, confirming that the structure of the atomistic water is preserved on passing from all-atom MD to H-AdResS MD.

**Water dynamics.** It is well known that the presence of a macromolecule in solution influences the diffusion of water in its vicinity [56]; anomalous diffusion processes to which the water molecules undergo close to the disordered surface of the biopolymer, have been observed both by simulations and experiments [77]. Hence, water dynamics may play an important role for drug binding[78]. On such a ground, we have computed the mean square displacement (MSD) of water (Figure 10) to verify whether our simulations are able to reproduce such diffusive properties.

In the log-log plot, the MSD trend of bulk water appears linear with time, representing normal diffusion. On the other hand, hydration water shows a clear deviation from linearity. In particular, two different intervals can be distinguished: one from 0.5 to 5 ps, where the hydration water diffusion cannot be described by a Brownian process, and one from 5 to 100 ps, where a linear dependence is recovered. Our results confirm that H-AdResS simulations correctly reproduce both normal diffusion in the bulk water and anomalous diffusion in the hydration shell, with respect to the fully atomistic simulations.

**Decreasing the atomistic region size.** We calculated the radius of gyration (Figure SI 7.a) and the number of intra-molecular hydrogen bonds (Figure SI 7.b) of the proteins in simulations with reduced $d_{at}$, relative to the values in fully atomistic simulations. We next calculated the error in the RMSF for the protein $C_\alpha$ atoms, relative to the RMSF in the fully atomistic system (Figure SI 7.c). Our analyses show that, for both proteins, the structures are correctly maintained and there are not significant changes in the local backbone translational flexibility in the range of $d_{at}$ considered. Finally, the calculated THz IR spectra (Figure SI 8) do not show significant differences with respect to the all-atom simulations.

As far as the water properties are concerned, the calculated water density (Figure SI 9) is uniform, apart from small deviations in the hybrid region, as in the case of the systems with larger $d_{at}$. The probability distributions of the tetrahedral order parameter for the hydration water molecules (Figure SI 10) are preserved in all the H-AdResS simulations. This suggests that water structural properties are maintained when decreasing $d_{at}$. However, the reorientational times of the water OH bonds (Figure 11) turn out to present increasing deviations from the all-atom simulations. Significant variations are observed when the difference between the thickness of the atomistic region and the radius of gyration of the protein (corresponding to ≈1.1 nm for atox1 and ≈1.4 nm for cyclophilin J) is equal or lower than 1.6 nm. Hence, caution should be taken when setting the value of $d_{at}$, as it may affect water dynamic properties. These results are in agreement with those obtained in Ref. [79], where a similar setup, based on the force-based AdResS method [38], was employed to investigate the properties of ubiquitin in water.

## 5. CONCLUSIONS

This paper has provided, for the first time, an example of how the Hamiltonian adaptive resolution simulation scheme can be employed in the study of proteins in the presence of a dual resolution solvent. In addition to the advantages of the force-based adaptive scheme so far implemented in the literature [38, 80], a Hamiltonian-based formulation such as that of H-AdResS actually represents a natural framework for the calculation of ligand binding energetics. The comparison with all-atom simulations shows that both structural and dynamical properties of the solvent and, more importantly, of the solute are correctly reproduced.

Using the same approach of ref. [79], we systematically reduced the radius $d_{at}$ of the spherical high-resolution domain where protein and solvent are treated at the fully atomistic level, while at the same time monitoring protein flexibility and vibrations as well as water structure and dynamics. All of them, but hydration water dynamics, correctly reproduced the all-atom MD in the range of $d_{at}$ simulated, chosen so as to avoid interactions between the atomistic protein and the fully coarse-grained water molecules. Hydration water dynamics is not well reproduced for $d_{at}$ lower than 2.5 in case of atox1 and 3.0 nm in case of cyclophilin J, suggesting that one should choose a value of $d_{at}$ that exceeds the protein's radius of gyration by at least 1.6 nm, in good agreement with previous data [79].

Finally, it's worth noting that our final goal is to couple an efficient dual resolution solvent based on H-AdResS with hybrid MM/CG models of membrane proteins. For these systems, standard bioinformatic approaches combined with fully atomistic simulations might fail due to the paucity of experimental structures and the low sequence identity. Within the planned implementation, atomistic water above the receptor cavity will be coupled to a coarse-grained reservoir through the H-AdResS scheme, while the protein region far from the binding region will be treated at CG level. This may pave the way to the use of open-boundary simulations to predict ligand affinity for their target membrane receptors.

**FIGURES**

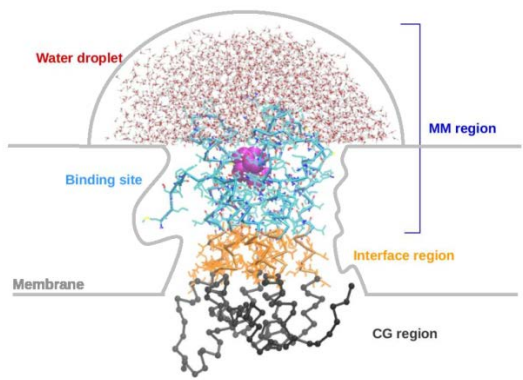

**Figure 1.** Representation of the regions described at different resolutions in the MM/CG approach. The figure shows a model of the GPCR TAS2R46 in complex with its agonist strychnine, investigated in ref. [20]. The hemispherical wall represents the boundary potential which prevents water evaporation.

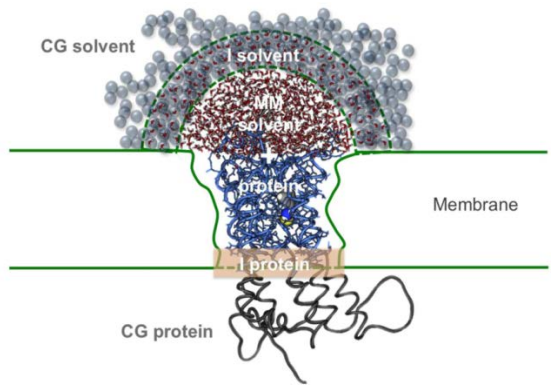

**Figure 2.** Scheme of the implementation of an adaptive multiscale approach on the MM/CG scheme. The water in the MM/CG upper hemisphere can freely diffuse out of the atomistic region, exchanging with the (coarse-grained) reservoir.

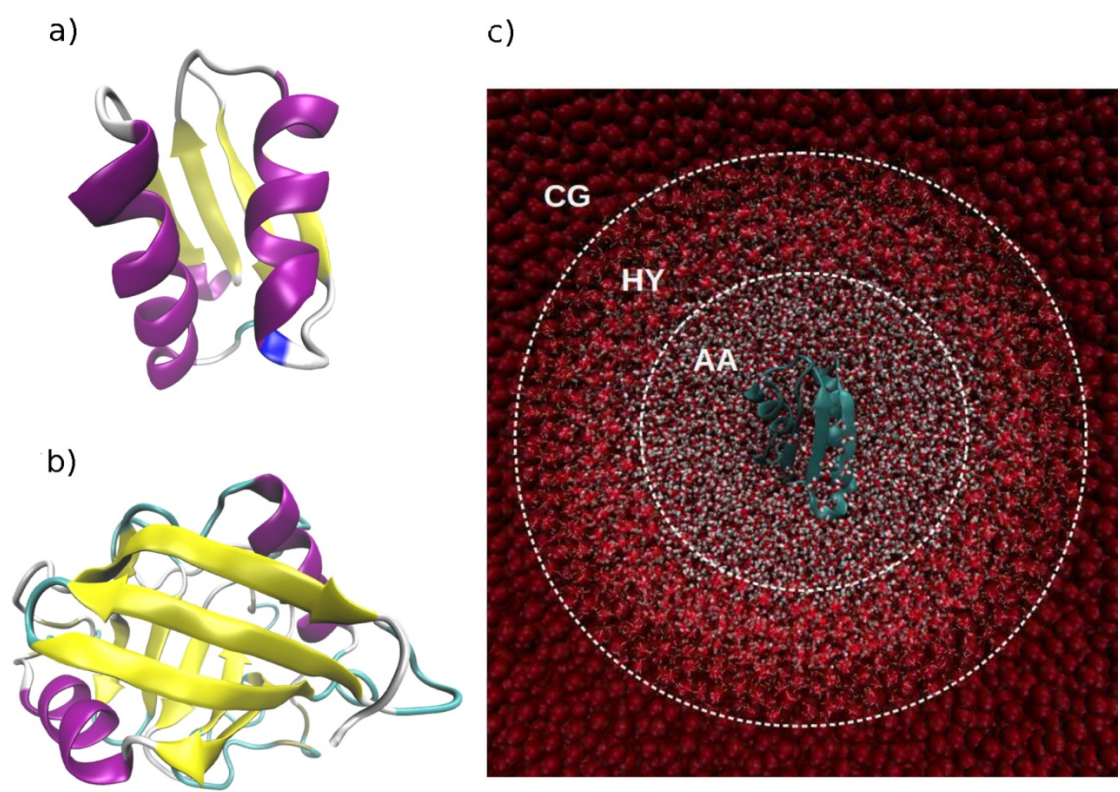

**Figure 3.** Ribbon representations of the Atox1 (**a**) and Ciclophilin J (**b**) proteins. (**c**) Set up of the H-AdResS simulation; in particular, the geometry employed for the atomistic (AA), hybrid (HY) and coarse-grained (CG) regions is shown.

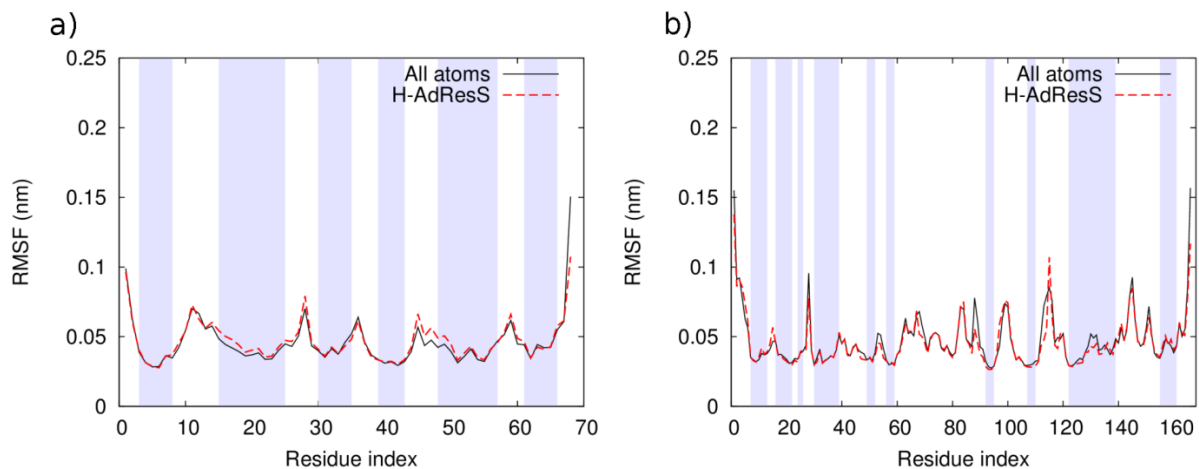

**Figure 4.** Root mean square fluctuations (RMSF) of the protein α-carbons, for the fully atomistic and H-AdResS simulations of atox1 (**a**) and cyclophilin J (**b**). The structured regions of the two sequence are highlighted.

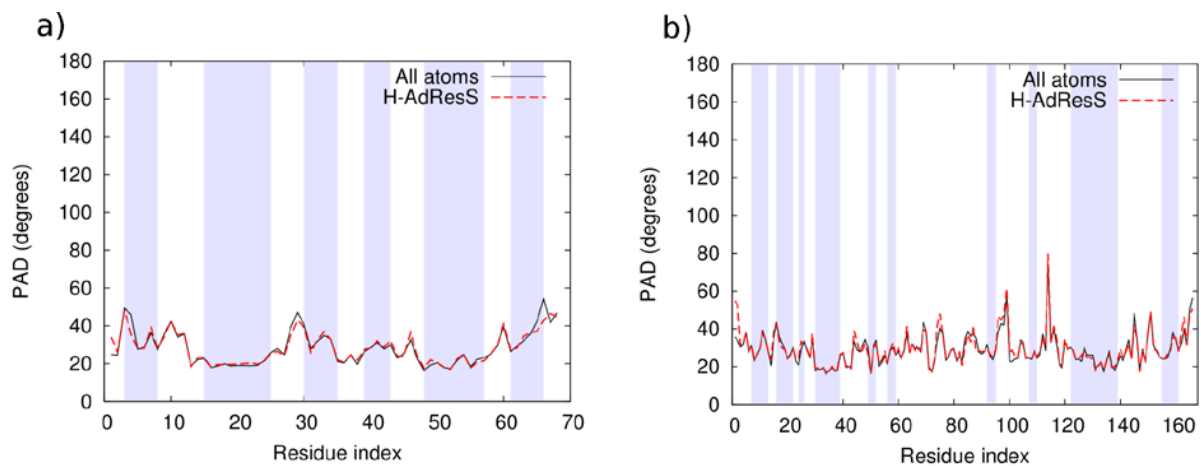

**Figure 5.** $PAD_\omega$ per residue, for atox1 (**a**) and cyclophilin J (**b**), in the fully atomistic and H-AdResS simulations. The structured regions of the two sequence are highlighted.

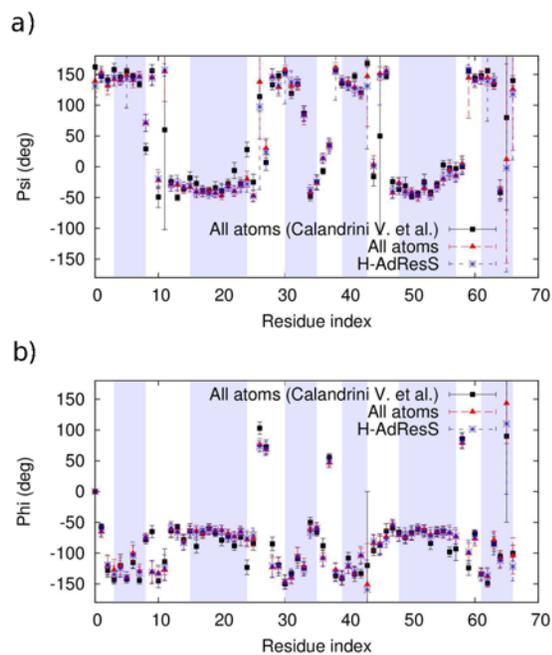

**Figure 6.** Average dihedral angles $\psi$ (**a**) and $\varphi$ (**b**) and standard deviation as a function of the residue number for the atox1 protein, in fully atomistic and H-AdResS simulations. Results from Ref. [73] are also reported for comparison.

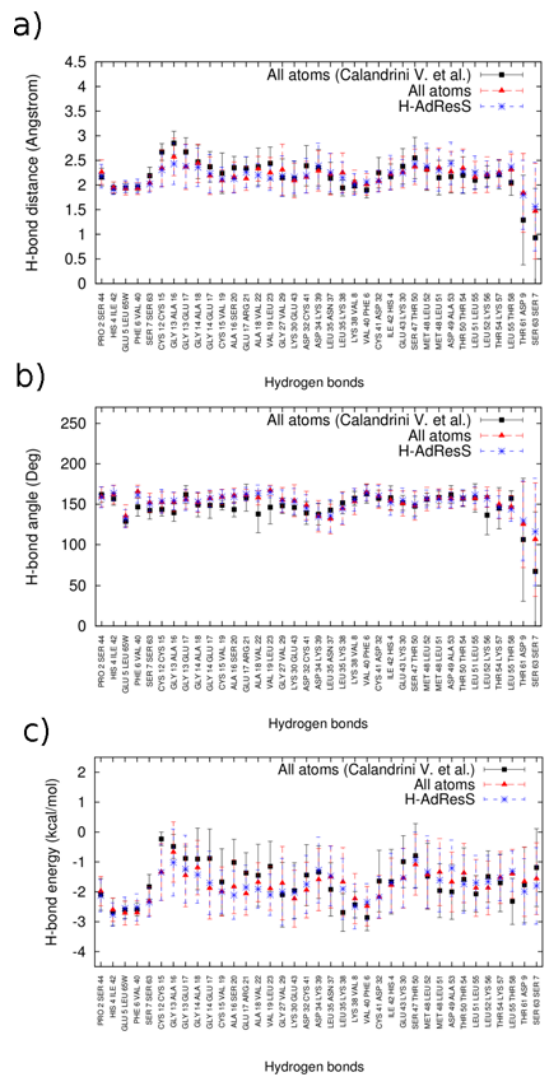

**Figure 7.** Average distance (**a**), angle (**b**) and energy (**c**) of the main-chain hydrogen bonds in atox1, in atomistic and H-AdResS simulations. Results from Ref. [73] are also reported for comparison.

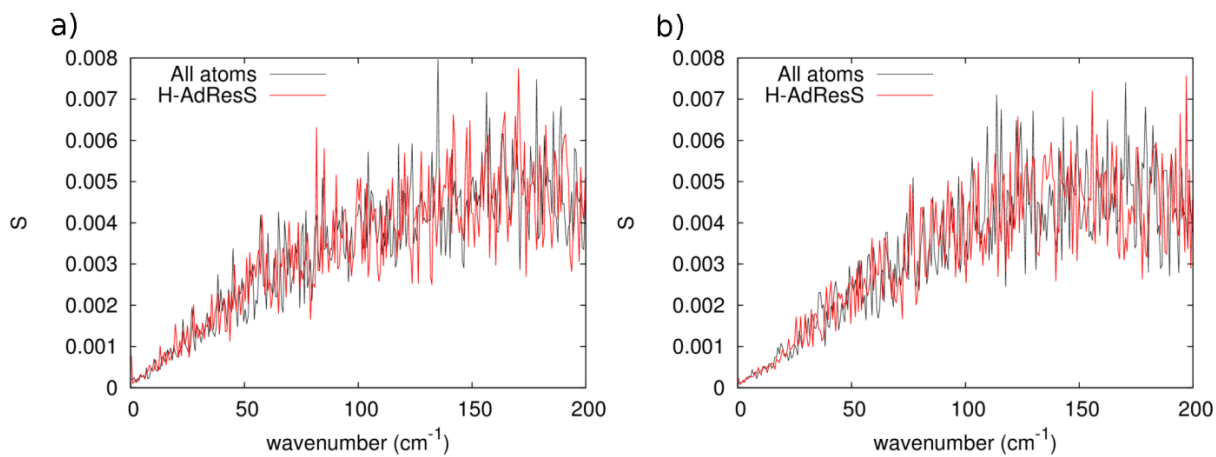

**Figure 8.** THz IR spectra of atox1 and cyclophilin J, respectively. The spectra have been calculated using VMD 1.9.2 with a superposition parameter of 20.

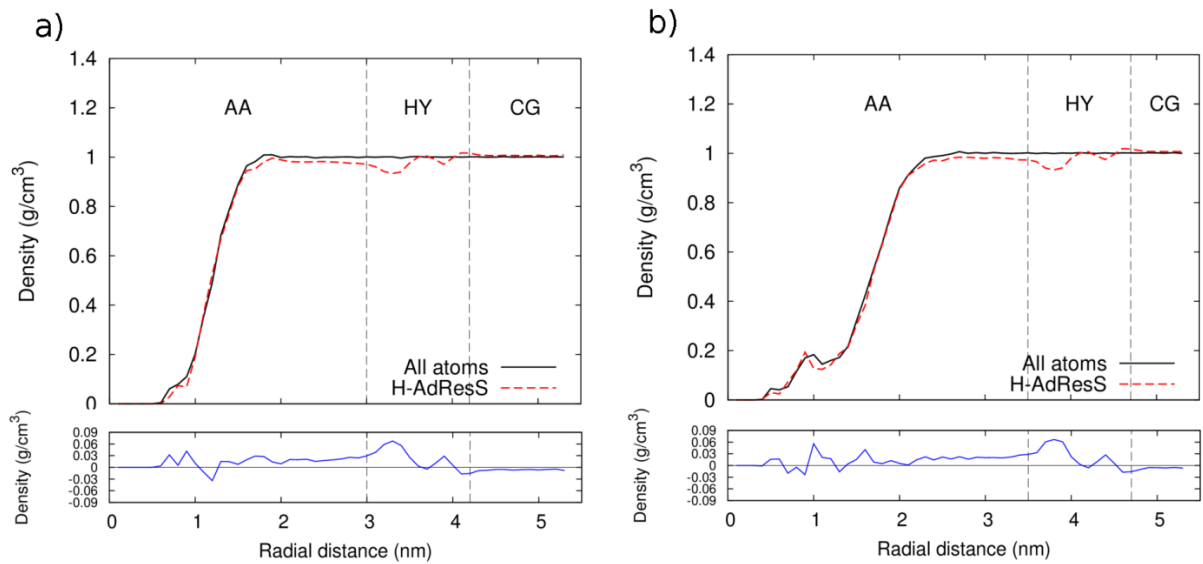

**Figure 9.** Radial density profile of water from the center of the atomistic region, across the atomistic (AA), hybrid (HY) and coarse-grained (CG) subdomains, for the atox1 (**a**) and cyclophilin J (**b**) systems. The residual plots are displayed in the lower panels.

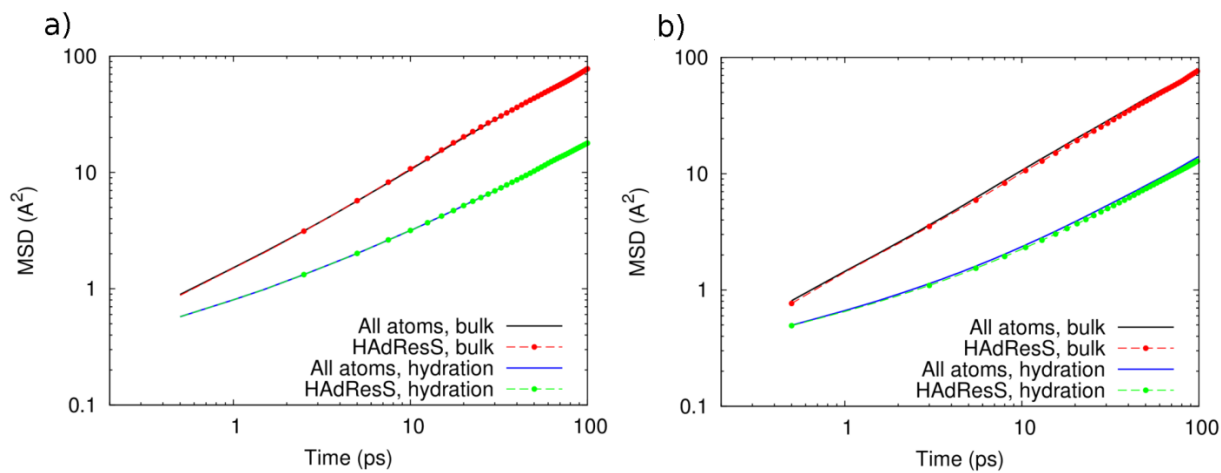

**Figure 10.** Mean square displacement (MSD) for hydration and atomistic bulk water in the atox1(**a**) and cyclophilin J (**b**) systems, for both the fully atomistic and H-AdResS simulations.

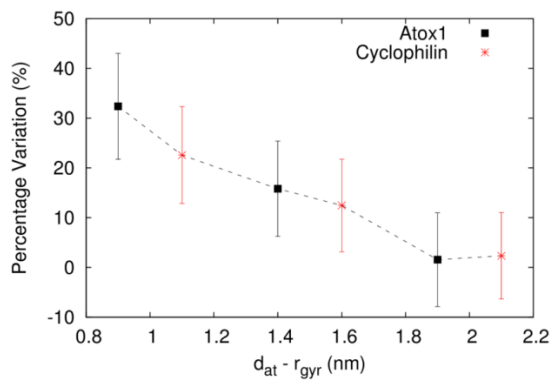

**Figure 11.** Percentage variation of the reorientational times of hydration water in H-AdResS simulations relative to all-atom ones, as a function of the distance between the radius of gyration and the border of the atomistic region. The error along the x-direction is within the point width.

**TABLES**

| Property computed | Simulation length | Sampling step |
|---|---|---|
| Structural properties (of water and protein) | 10 ns | 10 ps |
| THz IR spectra, reorientation tcf | 1 ns | 0.05 ps |

| Mean square displacement | 2 ns | 0.5 ps |

**Table 1.** Lengths and sampling time steps of the production runs used to calculate the structural and dynamical properties of water and protein.

| **Statistic** | (*Calandrini V. et al.*[73]) | **Fully atomistic** | **H-AdResS** |
|---|---|---|---|
| N° res with Gauche+ $\chi$ | 23.67+/-0.47 | 21.98+/-1.26 | 22.88+/-1.79 |
| N° res with Gauche- $\chi$ | 10.00+/-0.00 | 10.16+/-1.02 | 10.69+/-2.1 |
| N° res with Trans $\chi$ | 23.33+/-0.47 | 23.86+/-1.25 | 22.43+/-1.98 |
| Mean $\chi$ Gauche+ | -67.62+/-1.51 | -64.64+/-1.81 | -64.53+/-2.13 |
| Mean $\chi$ Gauche- | 60.21+/-1.59 | 63.92+/-2.88 | 63.56+/-3.27 |
| Mean $\chi$ Trans | 166.31+/-1.54 | 169.87+/-1.77 | 169.96+/-1.90 |

**Table 2.** Average number of residues with Gauche+ $\chi$, Gauche- $\chi$ and Trans $\chi$, and average values of $\chi$ Gauche+, $\chi$ Gauche- and $\chi$ Trans for the atox1 in atomistic and H-AdResS simulations. Results from Ref. [73] are also reported for comparison.

**AUTHOR INFORMATION**

**Corresponding Author**

* E-mail: v.calandrini@fz-juelich.de

**Notes**

The authors declare no competing financial interest.

**ACKNOWLEDGEMENTS**

This research is supported by the European Union's Horizon 2020-MSCA-ITN programme under grant agreement N° 642069 (HPC-LEAP project), which is funding the PhD position of one of the authors (TT). Simulations were performed with computing resources granted by JARA-HPC from RWTH Aachen University under project 10686. The authors are grateful to Kurt Kremer and Tristan Bereau for a careful reading of the manuscript and insightful comments.

## ASSOCIATED CONTENT

**Supporting Information.**

Derivation of the compensation term in the H-AdResS scheme; RMSD during equilibration phase; proteins' secondary structure as a function of time; solvent accessible surface area per residue; oxygen-oxygen and oxygen-hydrogen radial distribution functions of atomistic water; probability distributions of the tetrahedral order parameter for bulk atomistic and hydration water molecules; ratio between the radius of gyration in H-AdResS and fully atomistic simulations; relative number of intra-protein H-bonds in H-AdResS and fully atomistic simulations; error in the RMSF for the $C_\alpha$ atoms in H-AdResS and fully atomistic simulations; THz IR spectra of proteins for different widths of the atomistic region; radial density profile of water for different widths of the atomistic region; probability distributions of the tetrahedral order parameter for hydration water molecules, for different widths of the atomistic region. The supporting information is available free of charge via the Internet at http://pubs.acs.org.

**For Table of Contents use only**

**Title:** Open boundary simulations of proteins and their hydration shells by Hamiltonian adaptive resolution scheme.

**Authors:** Thomas Tarenzi, Vania Calandrini, Raffaello Potestio, Alejandro Giorgetti, Paolo Carloni.

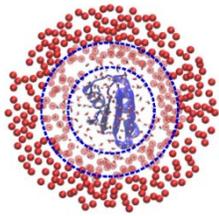

Supporting Information

**Open Boundary simulations of proteins and their hydration shells by Hamiltonian adaptive resolution scheme.**

Thomas Tarenzi, Vania Calandrini, Raffaello Potestio, Alejandro Giorgetti, Paolo Carloni

**SUPPORTING INFORMATION**

**SI 1. The origin of the term $\sum_{\alpha=1}^{N} \Delta H(\lambda_\alpha)$ in equation (1).**

The Hamiltonian in Eq. (4) in the main text can be rewritten as

$$
\begin{aligned}
H &= H_0 - \sum_{\alpha=1}^{N} \Delta H(\lambda_\alpha) \\
H_0 &= K + V^{p/p-w} + \sum_{\alpha}^{N} \left\{ \lambda_\alpha V_\alpha^{MM} + (1-\lambda_\alpha) V_\alpha^{CG} \right\},
\end{aligned}
\quad (1)
$$

where $V_\alpha^{MM}$ and $V_\alpha^{CG}$ are defined as:

$$
\begin{cases}
V_\alpha^{MM} \equiv \dfrac{1}{2} \sum_{\beta,\beta\neq\alpha}^{N} \sum_{i,j}^{n} V^{MM}\left(\left|\mathbf{r}_{\alpha i} - \mathbf{r}_{\beta j}\right|\right) \\
V_\alpha^{CG} \equiv \dfrac{1}{2} \sum_{\beta,\beta\neq\alpha}^{N} V^{CG}\left(\left|\mathbf{R}_\alpha - \mathbf{R}_\beta\right|\right)
\end{cases}
\quad (2)
$$

assuming only pairwise interactions.

A description of the system with the Hamiltonian $H_0$ in Eq. (1) leads to water density imbalance. The latter originates by the so-called drift force (arising from the coupling of the CG and MM Hamiltonians of the water molecules) and by the pressure imbalance between the CG and MM subsystems.

Supporting Information

The drift force appears when deriving the expression of the force acting on atom $i$ in water molecule $\alpha$, as obtained from $H_0$ in Eq. (1):

$$\mathbf{F}_{\alpha i} = \mathbf{F}_{\alpha i}^{p-w} + \sum_{\beta, \beta \neq \alpha}^{N} \left\{ \frac{\lambda_\alpha + \lambda_\beta}{2} \left( \sum_{j=1}^{n_\beta} -\frac{\partial}{\partial \mathbf{r}_{\alpha i}} V^{MM}\left(|\mathbf{r}_{\alpha i} - \mathbf{r}_{\beta j}|\right) \right) + \left(1 - \frac{\lambda_\alpha + \lambda_\beta}{2}\right) \left( -\frac{m_{\alpha i}}{M_\alpha} \frac{\partial}{\partial \mathbf{R}_\alpha} V^{CG}\left(|\mathbf{R}_\alpha - \mathbf{R}_\beta|\right) \right) \right\} \quad (3)$$
$$- \left[ V_\alpha^{MM} - V_\alpha^{CG} \right] \nabla_{\alpha i} \lambda_\alpha .$$

The first term in Eq. (3), $\mathbf{F}_{\alpha i}^{p-w}$, is due to the interactions with the protein. The second is a sum of pairwise forces obtained from MM and CG potentials, weighted by a function that is symmetric under molecule label exchange, that is $\alpha \leftrightarrow \beta$. The last term, $-\left[V_\alpha^{MM} - V_\alpha^{CG}\right]\nabla_{\alpha i}\lambda_\alpha$, is not antisymmetric under molecule label exchange. Hence, it breaks down Newton's Third Law in the hybrid region. This drift force pushes molecules into the all atom region where the Helmholtz free energy is locally lower, causing an imbalance of the density[1].

We remove the drift force by adding to $H_0$ a term (called here $\Delta H_{dr}$) whose form is determined by imposing that the drift force cancels on average, and by taking the integral of it:

$$\left. \frac{d\Delta H_{dr}(\lambda)}{d\lambda} \right|_{\lambda = \lambda_\alpha} = \left\langle \left[ V_\alpha^{MM} - V_\alpha^{CG} \right] \right\rangle_{\mathbf{R}_\alpha}$$
$$\Delta H_{dr}(\lambda) = \int_0^\lambda d\lambda' \frac{d\Delta H_{dr}(\lambda')}{d\lambda'} \quad (4)$$

The average $\left\langle \left[V_\alpha^{MM} - V_\alpha^{CG}\right] \right\rangle_{\mathbf{R}_\alpha}$ is performed by constraining the CG site of molecule $\alpha$ in the position $\mathbf{R}_\alpha$. It is here calculated approximately, under the reasonable assumption that the local environment of a molecule $\alpha$ having resolution $\lambda_\alpha$ is composed by other molecules with the same resolution:

$$\left\langle \left[V_\alpha^{MM} - V_\alpha^{CG}\right] \right\rangle_{\mathbf{R}_\alpha} \simeq \frac{1}{N} \left\langle \left[V^{MM} - V^{CG}\right] \right\rangle_{\lambda'} \quad (5)$$

---
[1] The drift force is proportional to the difference between the potential energies of a given molecule in the MM and CG representations.

Supporting Information

where $\lambda' \equiv \lambda(\mathbf{R}_\alpha)$ is the same for all water molecules. Therefore:

$$\Delta H_{dr}(\lambda) \simeq \frac{1}{N}\int_0^\lambda d\lambda' \left\langle \left[V^{MM} - V^{CG}\right]\right\rangle_{\lambda'} = \frac{\Delta F(\lambda)}{N} \qquad (6)$$

where $\Delta F(\lambda)$ is the Helmholtz free energy difference between a system with hybrid Hamiltonian at resolution $\lambda$ and the CG system, with $\lambda=0$. Therefore, one can calculate $\Delta H_{dr}(\lambda)$ by means of a simple Kirkwood thermodynamic integration [1]. Here, however, we employ an iterative scheme that corrects the drift force at each $\lambda$ value with the negative of its average value, following the approach defined in Ref. ([2]). This allows a more accurate determination of the compensation and a smoother coupling between the different subdomains of the setup.

This term, however, removes on average the drift force, but does not act on the density imbalance originating from the different virial pressures of the CG and MM subsystems [3]. In terms of the grand potential $pV$, this leads to $p_{MM}(\mu_{MM},T)V \neq p_{CG}(\mu_{CG},T)V$ (for identical volumes $V$), while the compressibilities remain unchanged [4] (having tuned the CG potential to match the radial distribution function of the atomistic water). In order to enforce a uniform density, a second compensating force can be introduced. It can be shown [4-5] that the integral of such force across the hybrid region, i.e. the work performed by this force on a molecule crossing the hybrid interface, can be approximated by the ratio between the local pressure profile and the reference (target) density. Hence, the approximate thermodynamic correction $\Delta H_{th}$, to be added to $H_0$, reads:

$$\Delta H_{th}(\lambda) = \frac{\Delta p(\lambda)}{\rho_0}. \qquad (7)$$

We specify that the performances of our current implementation of H-AdResS do not outdo yet those of the standard Gromacs release; this is explained by the extremely high degree of optimization of the officially released versions of the Gromacs software. Many of the standard Gromacs routines for the force calculation are replaced by new pieces of code in the H-AdResS simulation. Here, our aim is to give a proof of concept about the applicability of the H-AdResS scheme to biological macromolecules in solution; an optimization of the newly introduced routines is planned, before moving to more sophisticated applications.

Supporting Information

**Figure SI 2**

Root mean square deviation (RMSD) of backbone atoms of atox1 (**a.**) and cyclophilin J (**b.**) during equilibration.

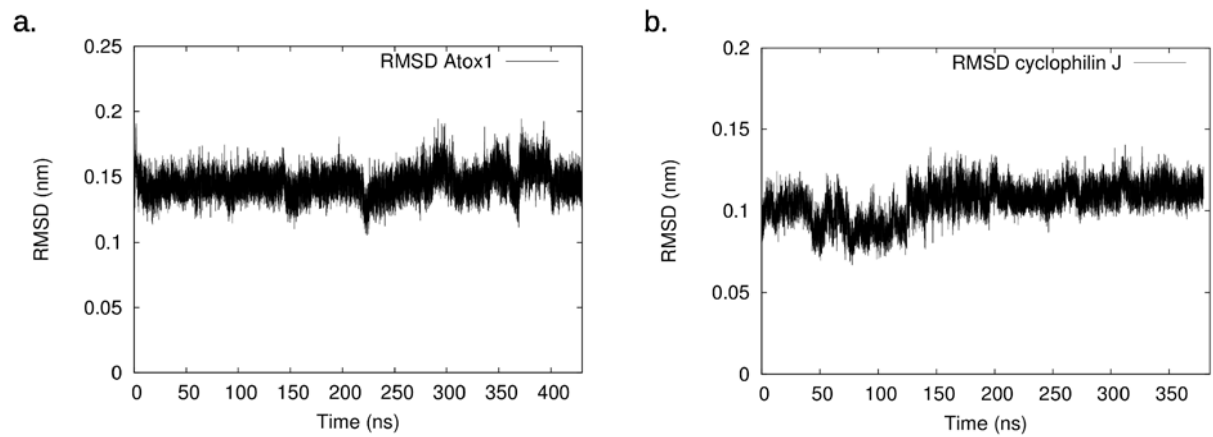

Supporting Information

**Figure SI 3**

Secondary structure for atox1 (**a.**, **b.**) and cyclophilin J (**c.**, **d.**) as a function of simulated time during the fully atomistic and H-AdResS simulations, respectively.

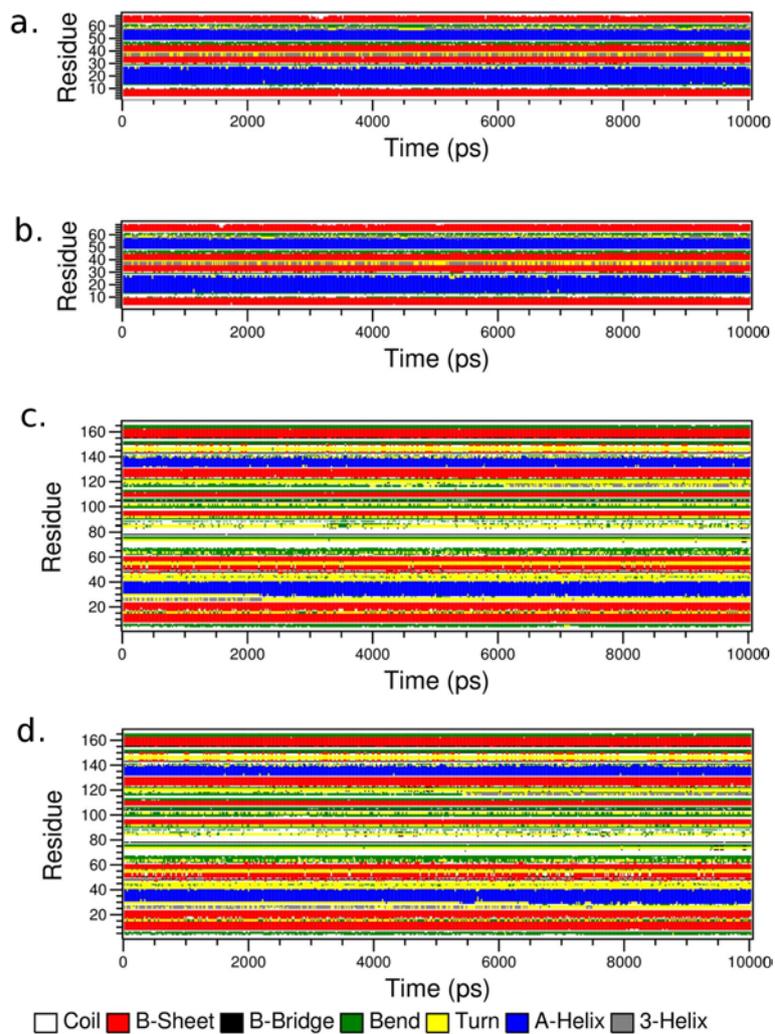

Supporting Information

**Figure SI 4**

Solvent accessible surface area (SASA) per residue, for atox1 (**a.**) and cyclophilin J (**b.**), in the fully atomistic and H-AdResS simulations. The residual plots are shown in the lower panels, displaying the differences between the fully atomistic and dual-resolution simulations.

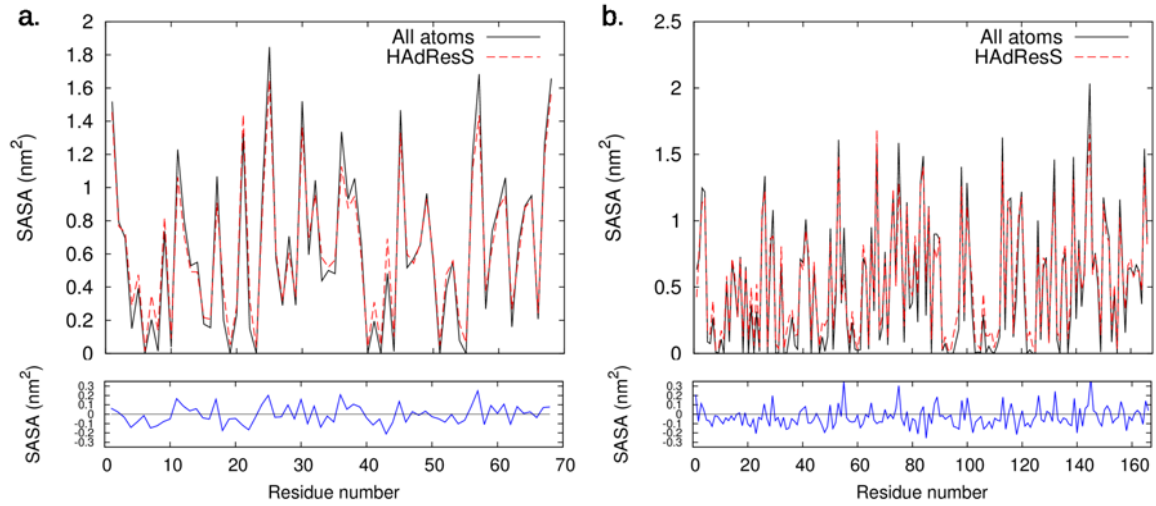

Supporting Information

**Figure SI 5**

Oxygen-oxygen (**a.**, **c.**) and oxygen-hydrogen (**b.**, **d.**) radial distribution functions (rdf's) of atomistic water solvating atox1 (**a.**, **b.**) and cyclophilin (**c.**, **d.**) as obtained by H-AdResS and fully atomistic simulations. Since our aim is just to compare the atomistic solvent in the H-AdResS simulation and the solvent in the corresponding region of the fully atomistic system, the rdf's are not corrected for the excluded volume due to the presence of the protein.

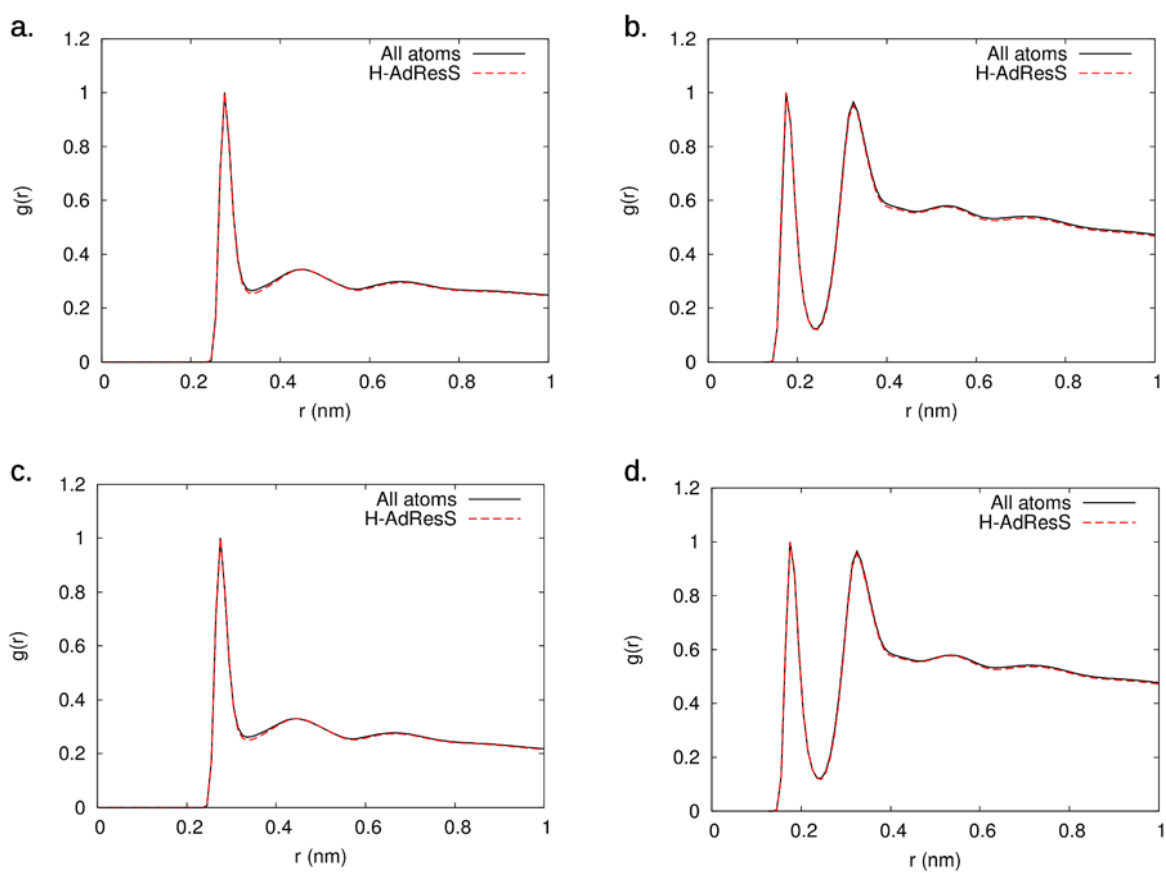

Supporting Information

**Figure SI 6**

Probability distributions of the tetrahedral order parameter for bulk atomistic and hydration water molecules in the atox1 systems (**a.** and **b.**, respectively) and in the cyclophilin J systems (**c.** and **d.**, respectively). Hydration water is defined as those water molecules whose oxygen atom was within a cutoff distance of 0.7 nm from the nearest protein heavy atom. The bulk region was defined by atomistic water molecules more than 0.7 nm from the protein surface.

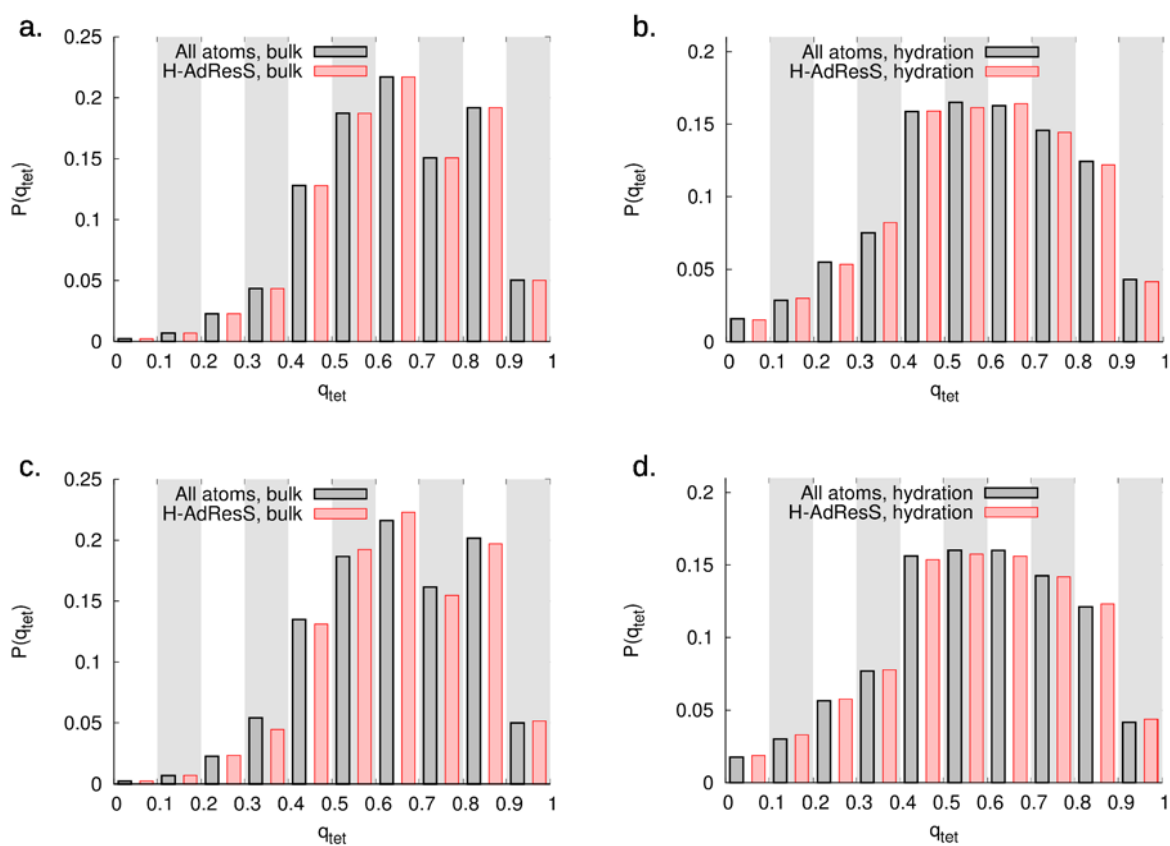

Supporting Information

**Figure SI 7**

**a.** Ratio between the radius of gyration of atox1 and cyclophilin J in the H-AdResS simulations (for different radii $d_{at}$ of the atomistic region) with respect to the fully atomistic ones, as a function of the distance between the radius of gyration of the protein and the border of the atomistic region. **b.** Relative number of intra-protein H-bonds in atox1 and cyclophilin J in the H-AdResS simulations with respect to the fully atomistic ones. **c.** Error in the RMSF for the $C_\alpha$ in atox1 and in cyclophilin J in the H-AdResS simulations with respect to the RMSF in the fully atomistic systems. In all the three plots, the error along the x-direction is within the point width.

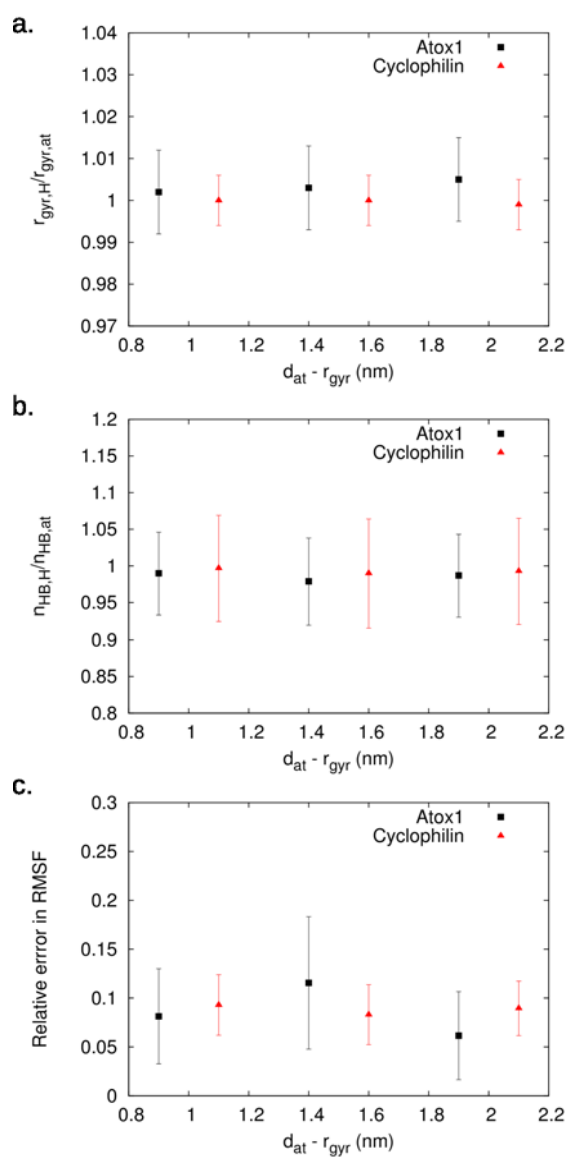

Supporting Information

**Figure SI 8**

THz IR spectra of atox1 (**a., b.**) and cyclophilin J (**c., d.**).The spectra have been calculated using VMD 1.9.2 with a superposition parameter of 20.

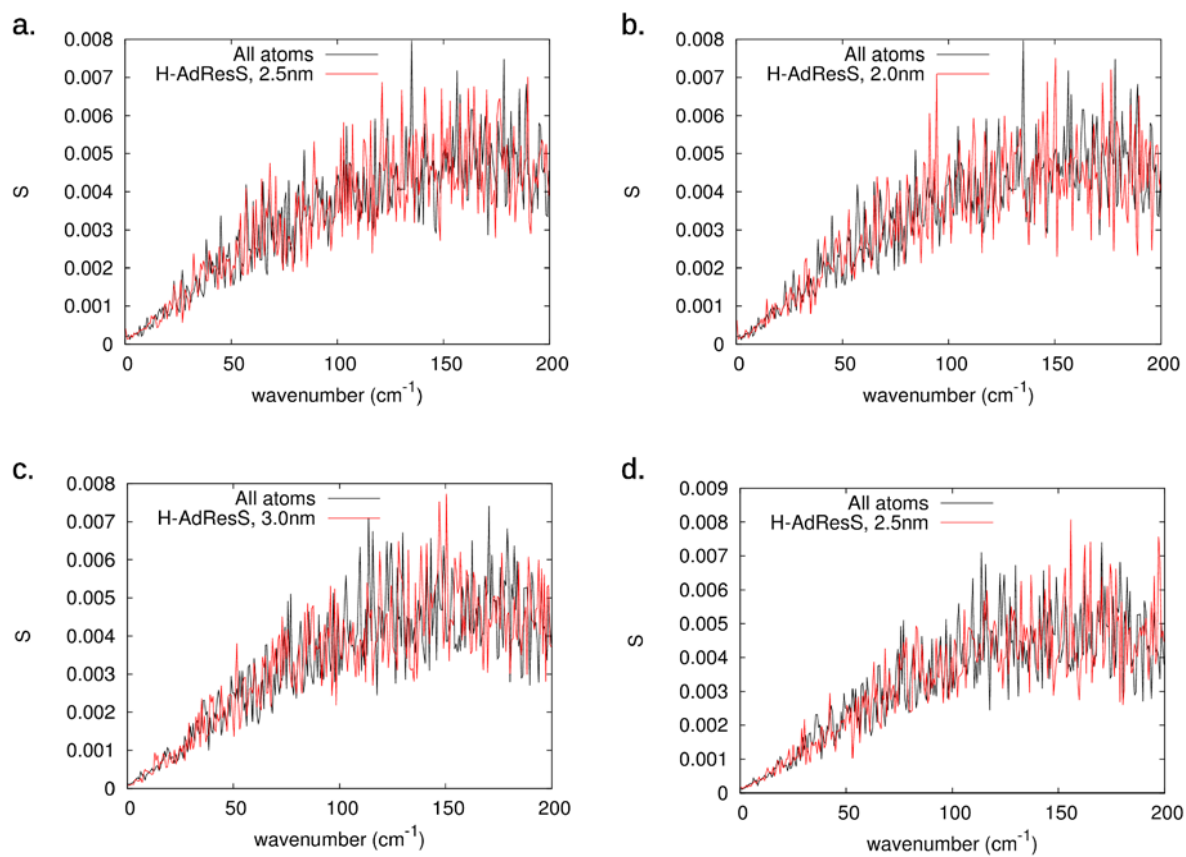

Supporting Information

# Figure SI 9

Radial density profile of water from the center of the atomistic region, across the atomistic (AA), hybrid (HY) and coarse-grained (CG) subdomains, for the atox1 (**a., b.**) and cyclophilin J (**c., d.**) systems.

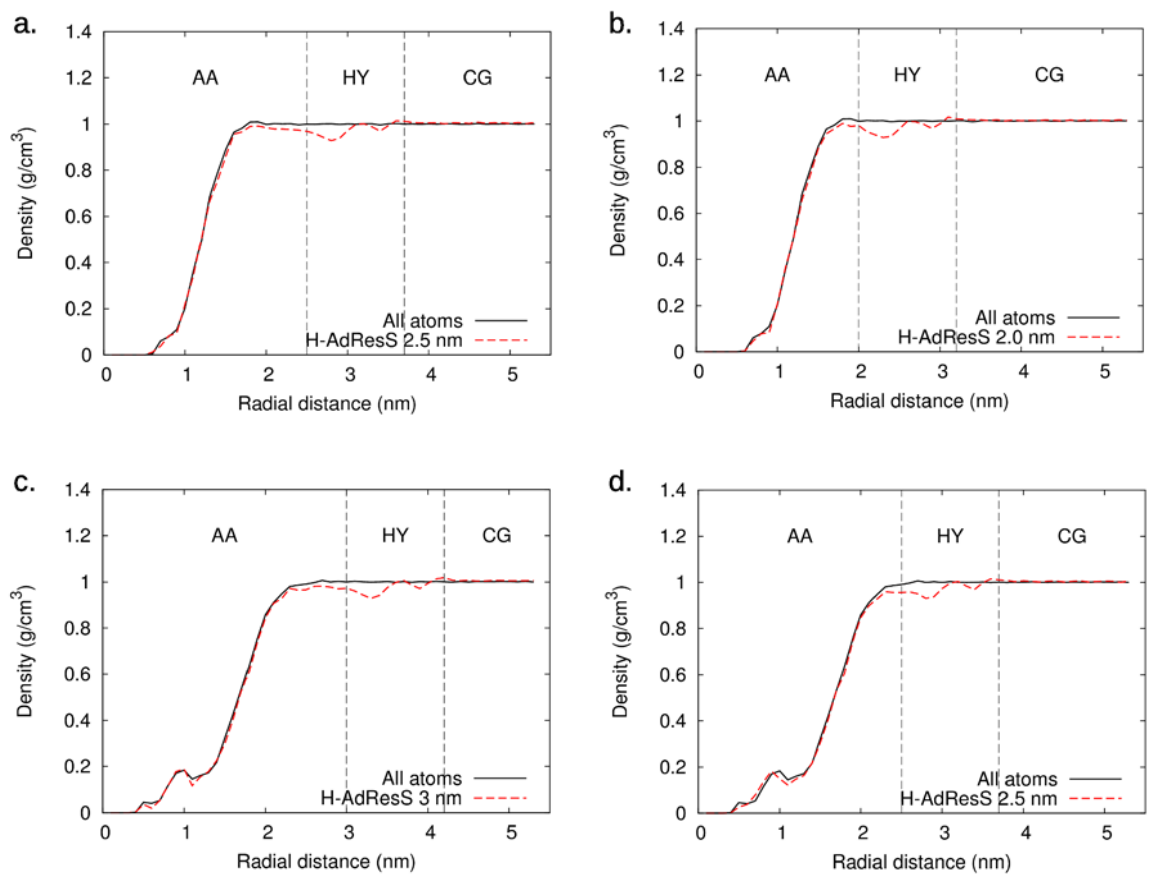

Supporting Information

# Figure SI 10

Probability distributions of the tetrahedral order parameter for hydration water molecules in the atox1 systems (**a.**, **b.**) and in the cyclophilin J systems (**c.**, **d.**).

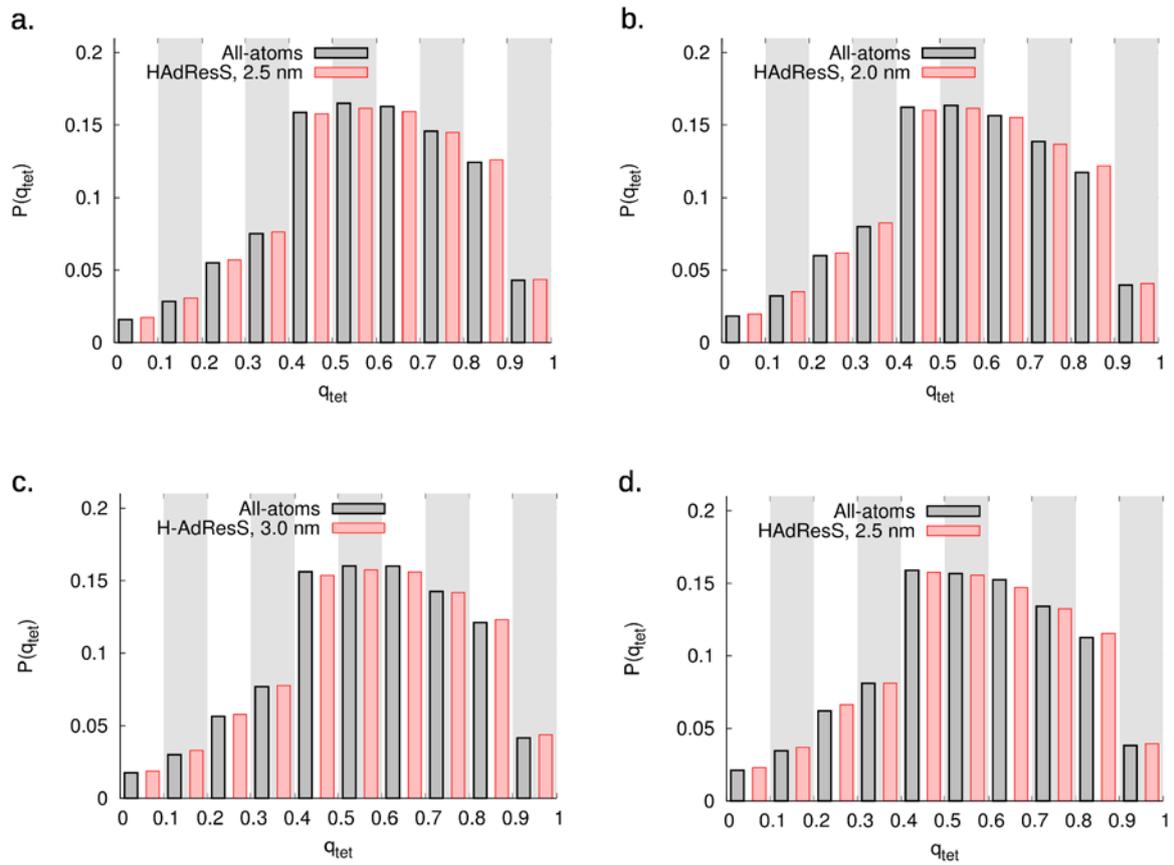

Supporting Information